\documentclass[twocolumn, notitlepage, amsmath, amssymb, superscriptaddress, floatfix]{revtex4-1}

\usepackage{graphicx}
\usepackage{dcolumn}
\usepackage{bm}
\usepackage{float}
\usepackage{braket}

\newcommand{\mywidth}{\linewidth}

\begin{document}

\title{\textsf{Continuous-wave room-temperature diamond maser}}

\author{Jonathan D. Breeze}
\email{jonathan.breeze@imperial.ac.uk}
\affiliation{Department of Materials, Imperial College London, Exhibition Road, London, SW7 2AZ, UK.}
\affiliation{London Centre for Nanotechnology, Imperial College London, Exhibition Road, London, SW7 2AZ, UK.}
\author{Juna Sathian}
\affiliation{Department of Materials, Imperial College London, Exhibition Road, London, SW7 2AZ, UK.}
\author{Enrico Salvadori}
\affiliation{Institute of Structural \& Molecular Biology, University College London, Gower Street, London, WC1E 8BT, UK.}
\affiliation{London Centre for Nanotechnology, 17-19 Gordon Street, London, WC1H 0AH, UK.}
\affiliation{School of Biological and Chemical Sciences, Queen Mary University of London, Mile End Road, London, E1 4NS, UK.}
\author{Neil McN. Alford}
\affiliation{Department of Materials, Imperial College London, Exhibition Road, London, SW7 2AZ, UK.}
\affiliation{London Centre for Nanotechnology, Imperial College London, Exhibition Road, London, SW7 2AZ, UK.}	
\author{Christopher W.M. Kay}
\affiliation{Institute of Structural \& Molecular Biology, University College London, Gower Street, London, WC1E 8BT, UK.}
\affiliation{London Centre for Nanotechnology, 17-19 Gordon Street, London, WC1H 0AH, UK.}
\affiliation{Department of Chemistry, University of Saarland, 66123 Saarbr\"ucken, Germany.}

\maketitle

\textbf{\textsf{The maser, older sibling of the laser, has been confined to relative obscurity due to its reliance on cryogenic refrigeration and high-vacuum systems. % A
Despite this it has found application in deep-space communications and radio astronomy due to its unparalleled performance as a low-noise amplifier and oscillator. % A
The recent demonstration of a room-temperature solid-state maser exploiting photo-excited triplet states in organic pentacene molecules \cite{Oxborrow, Breeze.15, Salvadori} paves the way for a new class of maser that could find applications in medicine, security and sensing, taking advantage of its sensitivity and low noise. % B
However, to date, only pulsed operation has been observed in this system.
Furthermore, organic maser molecules have poor thermal and mechanical properties, and their triplet sub-level decay rates make continuous emission challenging: alternative materials are therefore required. % B
Therefore, inorganic materials containing spin-defects such as diamond \cite{Poklonski.04, Poklonski.07, Jin.15} and silicon carbide \cite{Kraus.14} have been proposed.
Here we report a continuous-wave (CW) room-temperature maser oscillator using optically pumped nitrogen-vacancy (NV$^-$) defect centres in diamond. % D
This demonstration unlocks the potential of room-temperature solid-state masers for use in a new generation of microwave devices. %C
}}

Solid-state masers, developed in the 1960s, were realised by pumping paramagnetic impurity spin states, for example Cr$^{3+}$ ions doped into single-crystal sapphire (ruby).
Pumping three-level systems with microwaves generates a population inversion --- the requisite feature for amplification by stimulated emission in both masers and lasers, where a higher energy state is more populated than a lower energy state.

The first demonstration of a room-temperature solid-state maser required over 200 W of optical power to overcome the masing threshold, producing a burst of microwave power at 1.45 GHz.
This maser was subsequently improved and miniaturised using strontium titanate as the dielectric resonator material \cite{Breeze.15, Salvadori},  lowering the optical pump threshold and volume by two orders of magnitude.
However, the threshold pump rate per molecule remained the same at $\sim 10^4$ s$^{-1}$.
Hence, even if CW operation could be achieved in organic triplet-based masers, there is a severe drawback: thermal runaway, since almost all the absorbed optical pump power generates heat through non-radiative relaxation processes.
This presents a challenge since the thermal properties of organic hosts are typically poor.
For example, the pentacene host $p$-terphenyl has a very low thermal conductivity of 0.1 Wm$^{-1}$K$^{-1}$ and melting point of 230 $^\circ$C. 

Inorganic materials offer superior thermal and mechanical properties over organic materials.
Specifically, those hosting spin defects offer a potential route to realising CW room-temperature solid-state masers if the population of the defect spin states can be polarised (inverted).
Silicon carbide (SiC) is a promising candidate whose various types of spin-defect can be individually addressed by tuning the wavelength of optical excitation and externally applied magnetic field, which, through the Zeeman interaction, can select specific spin multiplicities.
V$_\textrm{Si}$ defects have spin-quadruplet ground-states which can be polarised through optical pumping.
Continuous stimulated emission of microwaves from optically pumped silicon vacancy (V$_\textrm{Si}$) defects in 6H-SiC has recently been reported \cite{Kraus.14}, but masing remained elusive.

In the 1960s, nitrogen impurities in diamond were proposed as potential maser gain media \cite{Smith.59, Sorokin.60, Siegman.64}, using a four-spin flip cross-relaxation mechanism to produce a population inversion in paramagnetic spin-$\tfrac{1}{2}$ nitrogen donors. 
Charged nitrogen-vacancy (NV$^-$) defect centres in diamond (see Fig. 1a) were observed using electron paramagnetic resonance (EPR) almost 40 years ago \cite{Loubser.78}.
The ability to prepare and read the quantum state of the triplet sub-levels efficiently using optically detected magnetic resonance has enabled many applications of NV$^-$ centres in magnetometry \cite{Gruber.97, Taylor.08, Maze.08} and quantum information processing \cite{Childress.06}.
They were recently proposed as quantum emitters for room-temperature masers \cite{Poklonski.04, Poklonski.07, Jin.15} due to their attractive properties of long spin dephasing times ($>$1 $\mu$s), long spin-polarisation lifetimes $\sim 5$ ms \cite{Takahashi.08, Jarmola.12} and triplet ground-states that can be polarized through optical pumping \cite{Robledo.11, Doherty.13}. %, Manson.06}.
Furthermore, diamond has the highest thermal conductivity $\sim 10^3$ Wm$^{-1}$K$^{-1}$ and excellent mechanical properties, which obviates thermal runaway.
Here we employ ensembles of optically pumped NV$^-$ centres as the gain medium for a CW room-temperature maser oscillator.

\begin{figure*}[ht]
\begin{center}
\includegraphics[width=\linewidth]{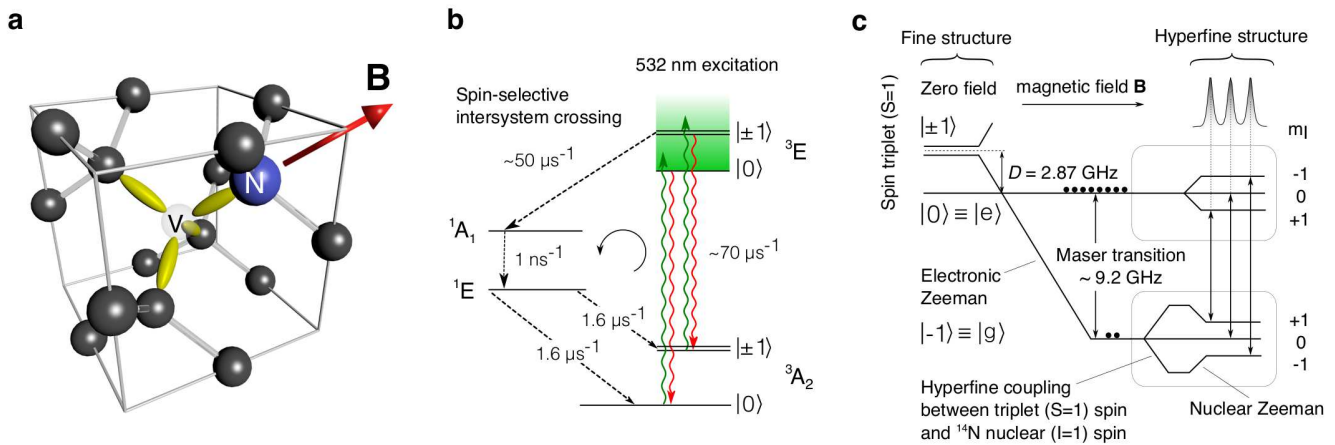}
\caption{
\textbf{a. NV$^-$ centre in diamond.} 
The NV centre comprises a nitrogen atom (blue) adjacent to a vacancy (light grey), surrounded by carbon atoms (black).  An external magnetic field $\mathbf{B}$ (red arrow) is applied along the N-V axis. 
\textbf{b. Optical spin-polarising pump process at zero-field.}
A continuous-wave 532 nm laser pumps electrons from the $^3A_2$ triplet ground-state into triplet excited-state $^3E$ via spin-conserving radiative transitions.
Non-radiative spin-selective intersystem crossing transfers roughly 40\% of the $\ket{\pm 1}$ electrons into singlet state $^1A_1$ which decay quickly into metastable state $^1E$, then back to the triplet ground-state with roughly equal probability of entering either $\ket{\pm 1}$ or $\ket{0}$ sub-levels.
The remaining electrons in the excited triplet state fluoresce back to the ground-state by spin-conserving radiative processes.
This process preferentially populates the $\ket{0}$ sub-level producing a spin-polarized triplet ground-state \cite{Robledo.11, Doherty.13}.
\textbf{c. Zeeman and hyperfine interactions of NV centres.}
A magnetic field $B$ applied parallel to the NV$^-$ axis, splits the $\ket{\pm 1}$ states through the Zeeman interaction.
For fields greater than $\sim 102.5$ mT the $\ket{-1}$ state energy drops below that of the $\ket{0}$ state.
If the $\ket{0}$ triplet sub-levels are preferentially populated through optical pumping, a population inversion is established.  Hyperfine interaction of ($S=1$) spin-triplets with adjacent $^{14}$N ($I=1$) nuclear spins produces three observable emission lines.
The maser transition is between the $\ket{e} \equiv \ket{0}$ and $\ket{g} \equiv \ket{-1}$ states. 
}
\vspace{7mm} 
\end{center}
\end{figure*}

\begin{figure}[ht]
\begin{center}
\includegraphics[width=\mywidth]{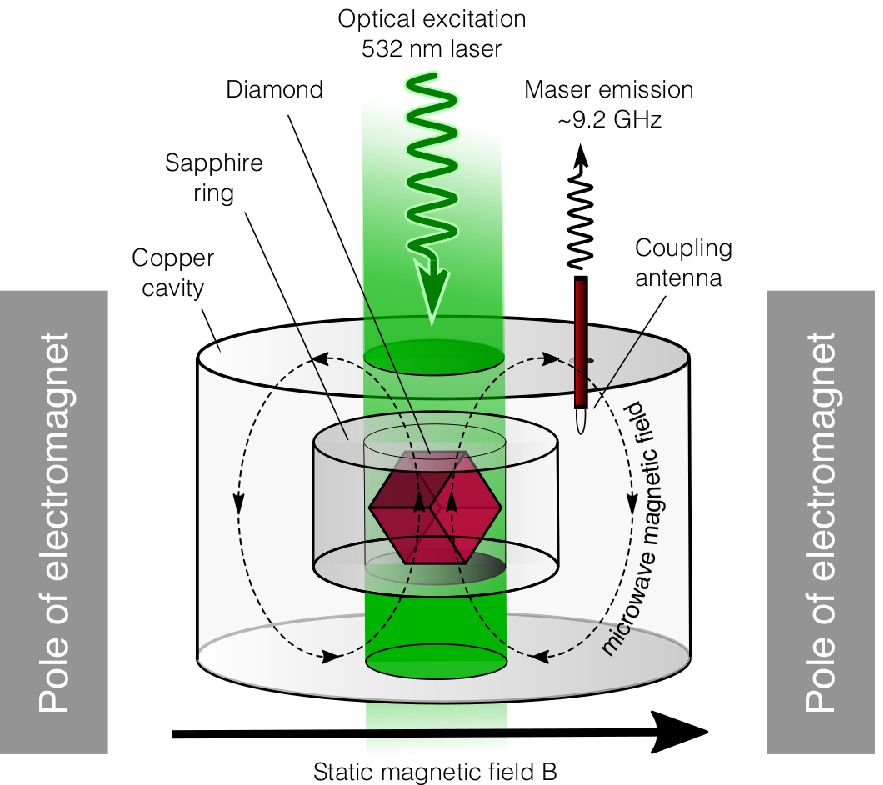}
\caption{\textbf{Diamond maser construction.}
The diamond sample is placed within a single-crystal sapphire ring such that a N-V axis ($[111]$ direction) lies within the plane perpendicular to the cylindrical axis of the ring.
A magnetic field B is applied across the resonator and the sample is rotated until it is aligned with the N-V defects.
The diamond is optically pumped at 532 nm by a continuous-wave Nd:YAG laser.
The sapphire and diamond are housed within a cylindrical copper cavity and a microwave loop antenna provides coupling to the outside world.  
}
\vspace{7mm} 
\end{center}
\end{figure}

The NV$^-$ defect centre has a triplet ground-state ($^3 A _2$) with associated zero-field splitting $D$ that places the quasi-degenerate $\ket{\pm 1}$ sub-levels approximately 2.87 GHz above the $\ket{0}$ sub-level (see Fig. 1b) \cite{Doherty.13}.
At room-temperature and in zero magnetic field, the sub-levels are populated according to Boltzmann statistics, with slightly more electrons populating the lower $\ket{0}$ sub-level than the $\ket{\pm 1}$ sub-levels.
The triplet ground-state can be spin-polarised by pumping with optical radiation with wavelengths in the region of 532 nm \cite{Robledo.11}, so that the $\ket{0}$ sub-levels become preferentially populated with respect to $\ket{\pm 1}$.
On photo-excitation, electrons in the $\ket{0}$ and $\ket{\pm 1}$ sub-levels of the triplet ground-state undergo spin-conserving transitions into an excited triplet state ($^3 E$).
Spin-selective intersystem crossing then preferentially transfers electrons from the $\ket{\pm 1}$ excited triplet-state sub-levels into a metastable singlet state, which then non-radiatively decay back to the ground-state, at roughly equal rates to the $\ket{0}$ and $\ket{\pm 1}$ sub-levels.
In this manner, continuous optical pumping of the NV$^-$ defect centre triplet ground-states can result in spin-polarisation of the triplet ground-state where up to 80\% of electrons reside in the $\ket{0}$ sub-level \cite{Doherty.13}.

\begin{figure}[ht]
\begin{center}
\includegraphics[width=\mywidth]{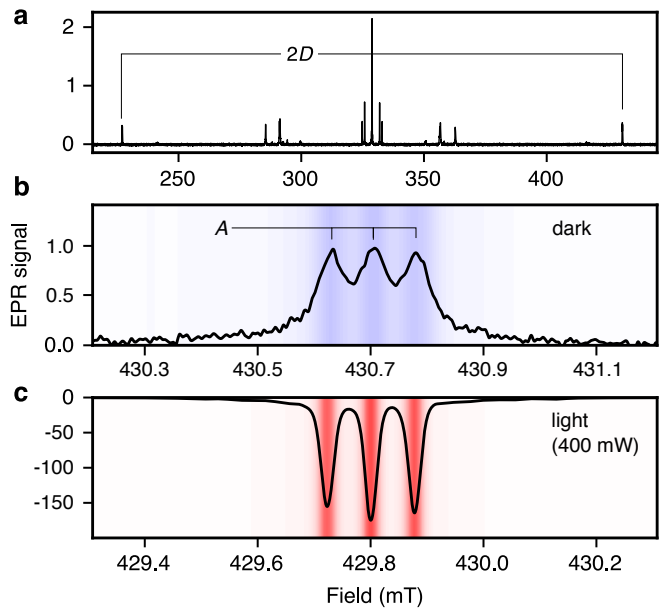}
\caption{\textbf{Electron paramagnetic resonance spectra.}
\textbf{(a)} Shows a wide EPR field sweep from 215 mT to 445 mT.  The diamond sample was aligned to provide maximum field splitting between the low-field and high-field lines.  \textbf{(b)} Shows the high-field absorption in the dark state, normalised to the maximum value.
The three lines are separated by $\approx 0.075$ mT, due to hyperfine coupling of ($S=1$) triplets with adjacent ($I=1$) $^{14}N$ nuclei. \textbf{(c)} The lines become emissive upon illumination by 400 mW of 532 nm laser light.
The EPR signal amplitude is $\sim 150$ times greater than the dark-state and has shifted -25 MHz in frequency ($\sim 0.9$ mT).
}
\vspace{7mm} 
\end{center}
\end{figure}

However, stimulated emission requires a positive population inversion.
This can be achieved by applying an external magnetic field along one of the four N-V defect axes (see Fig. 1a), splitting the energy of $\ket{\pm 1}$ states via the Zeeman interaction and leaving the energy of the $\ket{0}$ sub-level unchanged as shown in Fig. 1c.
For an electron Zeeman interaction energy, $\gamma_{\rm e} \hbar B$, greater than twice the zero-field-splitting energy $hD$ (\textit{i.e.} for $B \ge 102.5$ mT), the energy of the $\ket{-1}$ sub-level dips below that of the $\ket{0}$ state, permitting a population inversion to be established.
If the transition frequency $\omega_{\rm{s}}$ between the ground-state $\ket{g} \equiv \ket{-1}$ and excited state $\ket{e} \equiv \ket{0}$ is resonant with that of a cavity mode $\omega_{\rm{c}}$, then stimulated emission of microwave photons will occur.
Furthermore, if the rate of stimulated emission of microwave photons exceeds the loss-rate due to cavity dissipation then build up of the cavity photon population occurs, resulting in maser oscillation.

\begin{figure}[ht]
\begin{center}
\includegraphics[width=\mywidth]{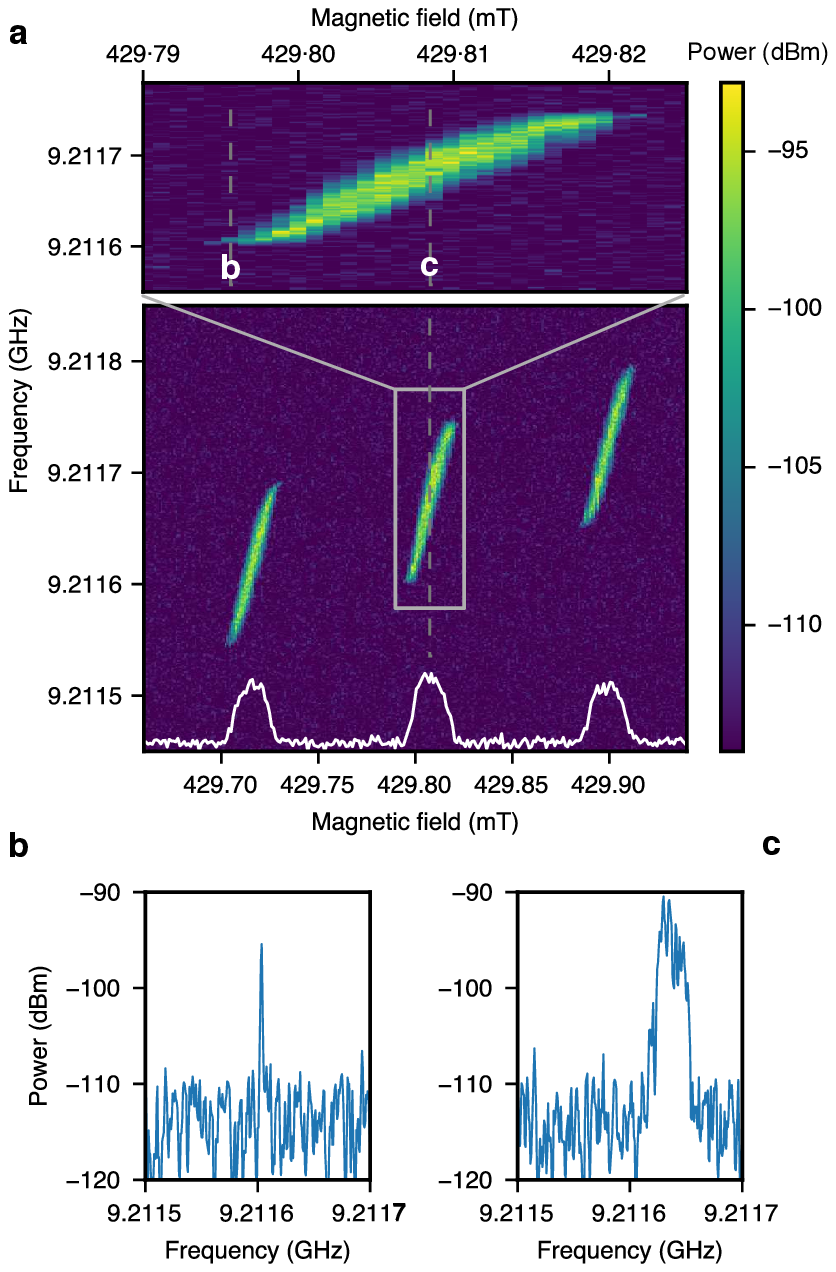}
\caption{\textbf{Field-frequency maser emission plots.}
For an optical pump power of 400 mW, the static external magnetic field was varied from 429 mT to 431 mT in 1 $\mu$T increments.
At each field value, the emission spectrum emerging from the maser was recorded.
\textbf{a}. This plot shows the maser emission spectrum as a function of externally applied magnetic field $\bm{B}$.  
There are three regions of maser oscillation corresponding to each of the hyperfine transitions.
The white line shows the integrated power spectral density.
The central region is magnified to show the narrow linewidth emission spectrum on the edge of threshold (\textbf{b}) and the limit-cycle broadened spectrum at the centre of the emission region (\textbf{c}).
\textbf{b}. At the edge of the emission spectrum, slightly above threshold the emission has narrow linewidth ($\sim 50$ Hz).  
\textbf{c}. At the centre of the emission region, the linewidth is broadened due to limit-cycles.
}
\vspace{7mm} 
\end{center}
\end{figure}

For this study, a synthetic type-IIa diamond doped with 1 ppm of nitrogen donors was grown by chemical vapour deposition (CVD), irradiated by high-energy electrons and annealed to produce 0.072 ppm of NV$^-$ centres (see Methods).
The diamond was laser cut into a cuboid of dimensions $2.1 \times 2.1 \times 2.6$ mm$^3$ with \{100\} faces and $\braket{111}$ crystal directions (along which the N-V defects lie) emerging from the corners of the cuboid.  
The number of NV$^-$ centres in the sample was estimated to be $1.5 \times 10^{14}$, of which one in twelve couple to the cavity mode due to the four N-V axes and three $^{14}N$ hyperfine lines, yielding $N = 1.2 \times 10^{13}$ active NV$^-$ centres.
The longitudinal (spin-lattice) and transverse (spin dephasing) relaxation times were measured at 9.5 GHz using pulsed electron paramagnetic resonance (EPR) spectroscopy yielding $T_1$ = 4.8 ms and $T_2^*$ = 20 $\mu$s respectively (see Methods).
These values are in agreement with those reported for samples with similar concentration of nitrogen \cite{Takahashi.08, Jarmola.12}.

The `cooperativity' $C = 4g_{\rm s}^2N/\kappa_{\rm c}\kappa_{\rm s}$ is a figure of merit for masers, where $g_{\rm s}$ is the single spin-photon coupling rate, $\kappa_c = \omega_c/Q$ the cavity mode decay rate, $Q$ is the cavity quality-factor and $\kappa_{\rm s} \approx 2/T_2^*$ the spin decoherence rate.
For above-threshold maser oscillation the cooperativity needs to be greater than unity, $C\gg 1$.
The Purcell factor is key to enhancing maser performance \cite{Breeze.15, Breeze.17} and contributes a factor $g_{\rm s}^2/\kappa_{\rm c}$ to the cooperativity.
A high Purcell factor maser cavity was therefore designed and constructed using a hollow cylindrical single-crystal sapphire dielectric resonator ($\varepsilon = 9.394$, outer diameter 10 mm, inner diameter 5.1 mm and height 6.5 mm) housed centrally within an oxygen-free high-conductivity copper cavity (diameter 36 mm and height 18 mm).
The cavity supports a TE$_{01\delta}$ mode resonating at 9.22 GHz with an unloaded $Q$-factor of 55,000.
A magnetic mode volume $V_m$ of 0.15 cm$^3$ was calculated, yielding a single spin-photon coupling of $g_{\rm s} = 0.2$ Hz and a $\sqrt{N}$-enhanced collective spin-photon coupling of $g_{\rm e} = 0.7$ MHz (see Methods).
The diamond was mounted inside the sapphire resonator and held in place by quartz tubes.
It was oriented such that two N-V $\braket{111}$ axes lay in the plane perpendicular to the cavity axis.
Coupling to the cavity field was achieved by a small adjustable antenna loop, set slightly below critical coupling ($k = 0.85$) to yield a loaded $Q$ of 30,000 and cavity mode decay rate of $\kappa_{\rm c} \approx 1.9$ MHz.
With a measured spin decoherence rate $\kappa_{\rm s} \approx 2/T_2^* = 0.1$ MHz, the cooperativity $C \approx 10.1$, indicating that the maser would oscillate above a threshold pump rate per active $NV^-$ centre of $w_{\rm{thr}}\sim 330$ s$^{-1}$ equivalent to 210 mW of optical power (see Methods for details).

The cavity was placed between the poles of an electromagnet (Bruker) as shown in Fig. 2 and oriented horizontally to allow optical access to continuous laser excitation (Laser Quantum Finesse Pure, $\lambda$ = 532 nm, spot size 2.25 mm, power $\le 10$ W) along the axis of the quartz tube.
The laser beam was adjusted until it was colinear with the cavity axis and $p$-polarized with respect to the surface of the diamond, since the angle of incidence was close to the Brewster angle ($67.2^\circ$).
The N-V defect axes were aligned with the magnetic field using CW EPR spectroscopy (see Methods).
EPR spectra were collected in the dark (see Figs. 3a, 3b) then laser excitation was applied with gradually increasing power.
As the laser power increases the high-field absorption lines decrease in amplitude until they disappear at a pump power of $\sim 1.5$ mW.
This is due to equalisation of the $\ket{e}$ and $\ket{g}$ populations, \textit{i.e.}, zero spin-polarization.
Modelling of the populations under optical pumping yielded an equalization pump rate $w_{\rm{eq}}$ of 1.5 s$^{-1}$, corresponding to a pump power of $\sim 0.9$ mW.
Considering Fresnel and coupling losses, this agrees fairly well with the measured incident power.
The low concentration of NV centres results in only 8.5\% of the incident optical power being absorbed.
Although this is far from optimal, it ensures the NV centres are pumped almost homogeneously, since the laser intensity varies marginally across the sample.

When the laser pump power was increased to 400 mW (see Fig. 3c), a cavity frequency shift of -25 MHz was observed.
This was attributed to an estimated 35 $^\circ$C increase in temperature of the sapphire ring and diamond, caused by heating.
The observed EPR signal was a factor $\sim 150$ greater in amplitude than the dark state.
Since the amplitude of the EPR signal is proportional to the difference in population of the $\ket{e}$ and $\ket{g}$ states (spin-polarization), it can be calibrated against the Boltzmann populated dark state to provide an estimate of the population inversion $S^z = N_{\rm e} - N_{\rm g} \approx 0.11 N$.
This agrees with the predicted inversion $S^z = \kappa_{\rm c} \kappa_{\rm s} / 4 g_{\rm s}^2$.
Once the optical beam alignment and polarization were optimized for maximum emission on the high-field line and the cavity frequency had stabilized, the maser was disconnected from the EPR spectrometer and connected to a spectrum analyzer (Advantest R3271A). 
The magnetic field was stepped across the high-field resonances using a programmable field controller (Bruker ER-031M) and the microwave output recorded at each magnetic field position.
Three separate regions of continuous maser emission were observed as shown in Fig. 4a, centred on 429.8 mT and separated by $\sim 0.075$ mT (hyperfine coupling), corresponding to the three emission lines observed with EPR.
The peak maser emission power was -90.3 dBm (comparable with a hydrogen maser) with an estimated pump rate per NV$^-$ centre of $w = 410$ s$^{-1}$.
The maser emission persisted without degradation in power for the duration of all experiments, the longest being 10 hours, demonstrating the robustness of the system.

On the edge of the maser emission spectra (see Fig. 4b), the linewidth $\sim 50$ Hz approaches the Schawlow-Townes limit \cite{Schawlow.58} for the observed output power: $
\gamma_{\rm{ST}} =  \tfrac{1}{2} \pi \hbar \omega_{\rm c} \kappa_{\rm c}^2/ P_{\rm{out}} \approx 10$ Hz.
However, near the centre of the each emission spectrum, the maser emission line appears broadened (see Fig. 4c).
The phenomenon of relaxation oscillation \cite{Siegman.86}, where the rate of stimulated emission is greater than the inversion replenishment rate, a normal occurrence when lasers or masers are switched on, usually decays with time.
Semi-classical dynamical simulations of the maser revealed oscillations that do not decay, instead exhibiting limit-cycles \cite{Dimer.07}.  
Further investigation revealed that they can be eliminated if the spin decay rate $\kappa_{\rm c}$ is increased (by increasing the concentration of NV centres) or if the spin-photon coupling rate $g_s$ is decreased through detuning ($\omega_{\rm c} \neq \omega_{\rm s}$), which explains why limit-cycles are absent just above threshold (see Fig. 4b).
% Cs atomic clock 9.193 GHz

The achievement of continuous maser operation resulted from an understanding that a high Purcell factor sapphire resonator was required in combination with the narrow linewidth of transitions in a diamond containing NV- defects. 
It should not be overlooked that although we used a frequency of 9.2 GHz in this study, which is close to the 9.193 GHz of a caesium atomic clock, any required frequency may be produced simply by changing the magnetic field and using a high Purcell factor cavity, resonant at the appropriate frequency. 
Hence, we expect that our work will form the basis for a new generation of microwave devices.

\section*{Methods}

\textbf{Spin Hamiltonian.}
The spin-Hamiltonian for NV$^-$ centres in diamond is given by \cite{Yavkin.16}:
\[
H_{\rm{spin}} = \gamma_{\rm e} \bm{B} \cdot \bm{S} - \gamma_{\rm n} \bm{B} \cdot \bm{I} + D \left[ S_z^2 - \tfrac{1}{3} S \left( S + 1 \right) \right] + \bm{S} \bm{A} \bm{I}
\]
where $\gamma_{\rm e}$ and $\gamma_{\rm n}$ are the gyromagnetic ratios of electrons and nuclei respectively, $D \approx 2.87$ GHz is the fine structure zero-field splitting, $\bm{S}$ and $\bm{I}$ are the triplet spin and $^{14}N$ nuclear spin eigenvectors, $\bm{A}$ is the uniaxially anistropic hyperfine coupling tensor ($A_\perp = -2.7$ MHz, $A_\parallel = -2.1$ MHz with respect to N-V axis) and $\bm{B}$ is the magnetic field.

\textbf{Spin-photon maser dynamics.}
The maser is modelled as an ensemble of $N$ two-level quantum emitters (lower state $\ket{g} \equiv \ket{-1}$ and upper state $\ket{e} \equiv \ket{0}$) resonantly coupled to a cavity mode.
The distance between the emitters is much less than the cavity mode wavelength and their spin-photon coupling is assumed to be homogeneous.
This interaction can be described by the Tavis-Cummings Hamiltonian \cite{Tavis.68} within the rotating wave approximation:
\begin{equation}
H_{\rm{TC}} = \hbar \omega_{\rm c} a^\dagger a + \frac{1}{2} \hbar \omega_{\rm s} S^z + \hbar g_{\rm e} \left( S^+ a + a^\dagger S^- \right),
\end{equation}
where $g_{\rm e} = g_{\rm s} \sqrt{N}$ is the enhanced collective spin-photon coupling strength, $g_{\rm s}$ is the single spin-photon coupling strength, $\omega_{\rm c}$ is the cavity frequency, $\omega_{\rm s}$ is the spin (two-level) transition frequency, $a^\dagger$ ($a$) are the cavity photon creation (annihilation) operators, $S^z$ is the collective inversion operator and $S^\pm$ are the normalized collective spin operators.
If the spin decoherence rate is much less than the cavity decay rate: $\kappa_{\rm s} \ll \kappa_{\rm c}$, then the single spin-photon coupling can be written \cite[Chapter~7]{Carmichael.03}:
\begin{equation}
g_{\rm{s}} = \gamma_{\rm e}\sqrt{\frac{\mu_0 \hbar \omega_{\rm c}}{2 V_{\rm m}} \frac{\kappa_{\rm s} / \kappa_{\rm c} }{1+\Delta^2}},
\end{equation}
where $\Delta = 2(\omega_{\rm c} - \omega_{\rm s})/\kappa_{\rm c}$ is a spin-cavity detuning parameter.
The Hamiltonian, coupled with a dissipative Liouvillian permits Lindblad master equations to be developed for the expectation values of the cavity field operator $a$, inversion $S^z$ and transverse spin $S^\pm$ operators \cite{Carmichael.03}.
These can can be solved in the steady state, leading to a simple maser threshold pumping rate per NV centre:
\begin{equation}
w_{\rm{thr}} = \gamma \left( C - 1 \right)^{-1} \eta
\end{equation}
where $C = 4g_{\rm s}^2N/\kappa_s \kappa_c$ is the cooperativity, $\gamma = 1/T_1$ is the spin-lattice relaxation rate and $\eta$ is a scaling factor based on the number of photons required to increase the inversion by one.
The scaling factor $\eta$ can be derived from the steady-state optical pump rate equations.
We derived a value of $\eta \approx 14.4$.
It holds for pump rates up to $w_{\rm{thr}}$ where the populations of the $\ket{\pm 1}$ states are approximately equal.
This is the maser amplification regime where the inversion, given by $S^z = wN/(w+\eta\gamma)$, is positive. 
Accessing this regime is not challenging since the dark-state populations of the triplet ground-states have approximately equal Boltzmann populations, only requiring weak optical pumping in order for the inversion to become positive.
One noticeable aspect of the threshold equation is that $w_{\rm{thr}}$ is lower than the scaled spin-lattice relaxation rate $\gamma\eta$ if $C>1$, which can be considered the condition for masing.
The inversion in the maser oscillation regime:
\begin{equation}
S^z = N_{\rm e} - N_{\rm g} = \kappa_s \kappa_c/4g_{\rm s}^2 = N/C 
\end{equation}
is always less than $N$ when $C>1$.
Above threshold and for stready-state conditions, we solve a non-linear system of rate equations comprising seven NV$^-$ states (triplet ground, triplet excited and singlet), a cavity field operator $\hat a$, a longitudinal (inversion) operator $\hat S^z$ and a transverse spin operator $\hat S^-$.
These equations are the Maxwell-Bloch equations coupled to a set of optical spin dynamical rate equations.

\textbf{Diamond preparation and characterisation.}
A synthetic diamond, doped with 1 ppm of nitrogen donors was grown by chemical vapour deposition.
The diamond was irradiated by high-energy electrons to create vacancies, then annealed to promote the formation of nitrogen-vacancy defect centres.
The diamond was laser cut into a cuboid of dimensions $2.1 \times 2.1 \times 2.6$ mm$^3$ with \{100\} faces and $\braket{111}$ crystal directions (along which the NV centres lie) emerging from the corners of the cuboid.  
These steps were performed by Element 6 (UK).
The diamond was cleaned in nitric acid to remove residual graphitic carbon (48 hours) then polished using a lint-free cloth.

Nitrogen-vacancy (NV) defect centres exist in neutral NV$^0$ and charged NV$^-$ states.
Using UV-Vis absorbance spectroscopy (Cary 5000 UV-VIS-NIR spectrophotometer, Ocean Optics USB 2000+ fiber optic spectrometer), the NV$^-$ concentration in the sample was found to be 0.072 ppm.
Further photoluminescence (PL) studies revealed that the ratio of charged to neutral NV centres (NV$^-$/NV$^0$) was $\approx 0.55$ under illumination from a 100 mW CW 532 nm Nd:YAG laser.
Furthermore, the ratio decreased slightly as the laser power was reduced, suggesting that some NV$^-$ had been photo-converted into NV$^0$ \cite{Manson.05, Aslam.13}. 
The total concentration of NV$^0$ and NV$^-$ was $\approx 0.2$ ppm.

\textbf{EPR spectroscopy of NV centres.}
X-band (9.5 GHz) pulsed EPR spectroscopy (Bruker Elexsys E580 spectrometer with ER 4118 X-MD5 resonator) coupled with pulsed laser excitation was employed to align the diamond sample and measure the relaxation times $T_1$ and $T_2^*$ of the high field line.
The diamond was held inside a hollow low-loss quartz tube and oriented inside the cylindrical single-crystal sapphire ring of the resonator so that two N-V axes lay in a plane perpendicular to the cylindrical axis.
The diamond was rotated about the cylindrical axis of the cavity until the external magnetic field was aligned with one of the two $\braket{111}$ directions and colinear with a N-V defect axis. 
This was achieved by sweeping the magnetic field from 200 mT to 450 mT and noting the minimum and maximum fields at which absorption/emission occurred. 
When the difference in magnetic field between the low-field and high-field lines was equal to twice the zero-field splitting $4\pi D/\gamma_{\rm e})$ or 205 mT, the absorption peaks associated with the other three $\braket{111}$  directions coalesced, becoming quasi-degenerate at $\sim 370$ mT.
Pronounced low-field absorption and high-field emission peaks appeared at 239 mT and 444 mT respectively. 
Nitrogen spin-$\tfrac{1}{2}$ (P1) impurities were evident at 341 mT.
The spin-lattice relaxation time ($T_1$) was measured using an inversion recovery sequence $\pi-T-\pi/2-\tau-\pi$-echo, where the interpulse delay $\tau$ was kept fixed at 1.2 $\mu$s and the time interval $T$ was incremented in steps of 10 $\mu$s from an initial value of 1 $\mu$s. 
The spin-dephasing relaxation time ($T_2^*$) was measured with a Hahn echo sequence $\pi/2-\tau-\pi$-echo, where the interpulse delay $\tau$ was incremented in steps of 80 ns from an initial value of $1.2$ $\mu$s. 
For both sequences the $\pi/2$ pulse length was 200 ns and each data set comprised 1024 points. 
The appropriate phase-cycling protocol was used in order to eliminate baseline offsets and contributions from unwanted echoes. 
The magnetic field was set on resonance with the $m_I = 0$ hyperfine line of the highest EPR band (see Figure 1c). 
Optical excitation was provided by nanosecond pulses from a $Q$-switched Nd:YAG laser (Continuum Minilite ML-II) with wavelength 532 nm, 4 ns pulse duration, 25 mJ pulse energy and 10 Hz repetition rate. 

\textbf{Maser cavity.}
The cavity was modelled using a radial mode-matching technique \cite{Kajfez.86}.
The electric filling factor of the sapphire was 0.76 and the magnetic filling factor of the central hole was 0.14.
The magnetic mode volume $V_{\rm m}$ was calculated from the ratio of the stored magnetic energy within the cavity to the maximum magnetic field energy density, $V_{\rm m} = \int_{\rm V} |\textbf{H}(\textbf{r})|^2 dV / |\textbf{H}(\textbf{r})|^2$.

\textbf{Optical pumping.}
The one-photon absorption cross-section of NV$^-$ centres at 532 nm is $\sigma = 3.1 \times 10^{-17}$ cm$^2$ \cite{Wee.07}.
The power required to pump the NV centres at rate $w$ is therefore $P_{\rm p} \approx w \hbar \omega_{\rm p} (A / \sigma)$, where $\omega_{\rm p}$ is the pump frequency and $A$ is the area (spot-size).
The defect pump rate $w = n\sigma I/\hbar\omega_{\rm p}$, where $n$ is the active NV concentration and and $\omega_p$ the optical pump frequency, is proportional to the local intensity $I = P_{\rm p}/A$ which varies across the sample.

% end of methods

\bibliographystyle{naturemag}

\bibliography{diamond-maser.bib}

\section*{Acknowledgements}

We thank Element 6 Ltd. (UK) for supplying the diamond.
We also thanks Paul French and Roy Taylor of the Photonics Group at Imperial College for generously lending us their CW laser.
This work was supported by the UK Engineering and Physical Sciences Research Council through grants EP/K011987/1 (IC) and EP/K011804/1 (UCL).
We also acknowledge support from the Henry Royce Institute.

\section*{Author Contributions}

J.B. conceived the study, developed the theory, designed the maser cavity, devised the experiment and wrote software for collecting experimental data.
J.S. characterised the diamond NV concentration by optical means and designed the optical pumping scheme.
J.B., C.W.M.K. E.S. and J.S. developed the experimental design and performed experiments.
J.B., C.W.M.K. and E.S. characterised the diamonds using EPR. 
J.B. interpreted the results with input from C.W.M.K. and E.S.
The paper was written by J.D.B., assisted by C.W.M.K. and with additional editing by E.S. and N.A. 

\end{document}